\documentstyle[prl,aps, multicol,epsf]{revtex}
\begin{document}
\draft
\title{Quasi-Long-Range Order in the Calogero-Sutherland Model } 
\author{Dora Izzo$^{*}$ and Gilson Carneiro}
\address{Instituto de F\'{\i}sica\\Universidade Federal do Rio de Janeiro\\  
C.P. 68528\\ 21945-970, Rio de Janeiro-RJ \\ Brazil }
\date{\today}
\maketitle
\begin{abstract}

The occurrence of quasi-long-range positional order in the
ground-state of the one-dimensional  repulsive Calogero-Sutherland
model is studied. By mapping  the exact ground-state into a one
dimensional classical system of interacting particles at finite
temperatures the structure function and the
displacement correlation functions  are calculated numerically 
using  Monte Carlo simulation methods. These are found to exhibit
quasi-long-range positional order for all values of the parameters.
The exponent characterizing the algebraic decay of the displacement
correlation functions with distance is estimated. 
It is argued that the ground-state of the repulsive Calogero-Sutherland
model consists of a single normal phase with quasi-long-range positional
order. 
\end{abstract}


\begin{multicols}{2}
\narrowtext

\section{Introduction} 
\label{sec.int}

It is well established that in a classical two-dimensional (2D) system
of particles, interacting through a potential that falls off
sufficiently fast with distance, long-range positional order cannot
exist at finite temperatures \cite{mer}. Instead, a low temperature phase
with quasi-long-range positional order (QLRPO) occurs. This  phase is
characterized by particle displacement correlation functions that  
decay algebraically with distance, with a temperature-dependent
exponent. This leads to  a structure function with power law singularities
at some reciprocal lattice vectors. It is predicted that this phase is
destroyed, at a temperature $T_c$, by a phase transition that can be
either first order or continuous.  The latter resulting from 
the unbiding of topological excitations. 
It is also known that theoretical considerations that take into
account only thermal phonon fluctuations in the harmonic approximation
correctly predict the low temperature properties of this phase \cite{nel}.
These predictions are confirmed by Monte Carlo simulations of
model systems \cite{tch}.

For a quantum system of bosons in one-dimension (1D), interacting
through a potential that falls off sufficiently fast at large distances,
long-range positional order is also ruled out in the ground-state.
For this system, theoretical analysis that accounts for  zero-point
phonon fluctuations in the harmonic approximation make predictions
analogous to those for classical 2D systems: destruction of long-range
positional order, power-law decay with distance of displacement correlation
functions and  power-law singularities in the structure function at some
reciprocal lattice vectors. However, QLRPO in the ground-state of such
1D bosons is not as well established as in  classical 2D system,

In this paper we study in detail positional order in a  1D
system of bosons interacting through a  repulsive potential that varies
inversely with the square of the distance - the Calogero-Sutherland
(CS) model -   for which the  exact ground-state wavefunction is known
\cite{cs}. 
The structure function for the CS model was calculated in the harmonic
approximation  by several authors\cite{strh}. The results confirm  that it
has power law singularities at reciprocal lattice vectors with an
exponent that varies continuously with the model parameters. However, 
the range of validity of this approximation it is not known. It is
unclear whether QLRPO persists outside this range and, if so, what is the
exponent governing the power-law decay of the displacement correlations. We
address these questions here. 

In order to go beyond  the  harmonic approximation we  resort to
numerical calculations. Sutherland obtained several years ago the exact
ground-state wavefunction  for the CS model \cite{cs}. Based on this
wavefunction we show that the ground-state of the CS model can be mapped 
into a  classical 1D system of interacting particles at finite
temperatures. By applying Monte Carlo (MC) simulation methods
\cite{sma} to this
classical system we calculate the structure function and investigate
the occurrence of QLRPO for arbitrary values of the model parameters. 

We show evidence that the CS model ground-sate has a single phase with
QLRPO and  argue that this phase is non-superfluid.
Over a range of parameter values we find that QLRPO 
leads to  power law singularities in the structure function and     
estimate the exponents governing these  singularities by
finite size scalling analysis of the MC data. Outside this range we find
that the structure function has no singularity, and that the displacement
correlation functions decay algebraically with distance. We also estimate
the exponent governing this decay. We find that this exponent differs
from  that predicted by the harmonic approximation  only in the
region were quantum fluctuations are large, being smaller than the
harmonic ones there. 

For particular values of the CS model parameters  the structure
function  has been calculated  exactly by Sutherland \cite{cs}. Our numerical
results are found to be in good agreement with the exact ones.

This paper is organized as follows. In Sec.\ref{sec.gsw} we review the exact
results obtained by Sutherland  and derive the mapping into a classical 1D
 interacting system. In Sec.\ref{sec.hap}  we review  the harmonic
approximation for the CS model and show that Sutherland's  exact
ground-state wavefunction reduces to the harmonic one in the
semiclassical region. In Sec.\ref{sec.nrs} we explain our numerical
method in detail and report the results  obtained by it. In
Sec.\ref{sec.dis} we interpret  these results and 
state our   conclusions.

\section{ Ground-State Wavefunction}
\label{sec.gsw}

The CS model describes $N$ bosons of mass $m$
on a line of length $L$ interacting through the two-body potential 
\begin{equation}
V(r)= \gamma \sum^{\infty}_{n=-\infty}
(r+nL)^{-2}=\frac{\gamma\pi^2}{L^2}[\sin{\frac{\pi r}{L}}]^{-2}
\; , \label{eq.vr} 
\end{equation}
where $\gamma $ is a constant that we assume positive. This potential
is periodic, with period $L$. For  large $L$  the $n=0$ term in Eq.\ (\ref{eq.vr}) is 
the most  important, so that $V(r)$ varies essentially as $r^{-2}$ \cite{cs}.

Sutherland found that the exact ground-state wavefunction for this
system is given by
\begin{equation}
\Psi= {\mbox { const.}} \prod_{i>j}\mid
\sin{\frac{\pi(x_i-x_j)}{L}}\mid^{\lambda} \; ,
\label{eq.gs}
\end{equation}
where 
\begin{equation}
2\lambda\equiv \beta = 1+(1+2g)^{1/2} \; . 
\label{eq.lb}
\end{equation}
The dimensionless parameter
$ g=\frac{4m\gamma}{\hbar^2} $
measures the relative strengths of the potential and kinetic energies.
Sutherland also found that the exact ground-state energy is given by
\begin{equation}
E=N\frac{\hbar^2}{2ma^2}\frac{\pi^2 \beta^2}{12} \; , 
\label{eq.egs}
\end{equation}
where $a=L/N$ is the mean interparticle distance.

The ground-state average of any operator that depends only on the 
position operators $x_i$ ($i=1,2,...,N$), $A(\{x\})$, is given by
\begin{equation}
\langle A \rangle =\frac{\prod^{N}_{i=1}\int^{L}_{0}dx_i A(\{x\}) \mid
\Psi \mid^2}{\prod^{N}_{i=1}\int^{L}_{0}dx_i  \mid \Psi \mid^2}
\; . 
\label{eq.avg}
\end{equation}
Using Eqs.\ (\ref{eq.gs}) and (\ref{eq.lb}) $\mid \Psi \mid^2$ can be
written as  
\begin{equation}
\mid \Psi \mid^2=\exp{\{-\beta \frac{1}{2}\sum_{i\neq j}{\cal V}(x_i-x_j)\}}
\; , 
\label{eq.ps2}
\end{equation}
where 
\begin{equation}
{\cal V}(x)=-\ln{\mid \sin{\frac{\pi x}{L}}\mid} \; .
\label{eq.vcl}
\end{equation}
Thus,
\begin{equation}
\langle A \rangle =\frac{\prod^{N}_{i=1}\int^{L}_{0}dx_i
A(\{x\})\exp{\{-\beta \frac{1}{2}\sum_{i\neq j}}{\cal V}(x_i-x_j)\}} 
{\prod^{N}_{i=1}\int^{L}_{0}dx_i \exp{\{-\beta \frac{1}{2}\sum_{i\neq
j}{\cal V}(x_i-x_j)\}}} \; ,
\label{eq.avt}
\end{equation}
According to this equation $\langle A\rangle$ may be calculated as the
canonical ensemble average of  $A(\{x\})$
of a 1D classical system of particles interacting through the
fictitious two-body
potential ${\cal V}(x)$. The parameter $\beta$ plays the role of the
inverse temperature.

The numerical calculations carried out in this
paper are based on Eq.\ (\ref{eq.avt}), as discussed in Sec.\ref{sec.nrs}

\section{ Harmonic Approximation}
\label{sec.hap}

Here we review the harmonic approximation  for the 1D boson
system described in Sec.\ref{sec.gsw}. 

\subsection{Phonon Spectrum and Wavefunction}
\label{sec.hps}

The interaction energy of the 1D boson system under consideration here is 
\begin{equation}
U=\frac{1}{2}\sum_{i\neq j} V(x_i-x_j)
\; . 
\label{eq.uvx}
\end{equation}
As usual, the harmonic approximation consists in expanding $U$ to second
order in the displacements, $u_n$, from the classical equilibrium
positions $X_n=na$ \cite{dp}. The result is
\begin{equation}
U=\frac{1}{2}\sum_{n\neq m} V[X_n-X_m -(u_n-u_m)]\equiv E_{cl} + \delta U
\; , 
\label{eq.udu}
\end{equation}
where $E_{cl}$ is the classical ground-state energy and $\delta U$ is the
harmonic correction 
\begin{equation}
\delta U=\frac{1}{2}\sum_{n\neq m}\frac{\partial^2 V(x)}{\partial
x^2}\mid_{x=X_n-X_m} (u_n-u_m)^2
\; . 
\label{eq.dud}
\end{equation}
In terms of Fourier transforms
\begin{equation}
\delta U = \frac{1}{2}\sum_k m \omega^2(k)\mid u_k \mid^2
\; , 
\label{eq.uhk}
\end{equation}
where $u_k$ is the Fourier transform of the displacement $u_n$,
$-\pi/a<k<\pi/a$, and the  phonon spectrum is given by
\begin{equation}
m\omega^2(k) = \sum_{n\neq 0}(1-e^{-ikX_n})\frac{\partial^2 V(x)}{\partial
x^2}\mid_{x=X_n}
\; . 
\label{eq.phs}
\end{equation}
Approximating $V(x)$ by $V(x)=\gamma/x^2$, the sums in Eq.\
(\ref{eq.phs}) can be performed exactly. The result is
\begin{equation}
\omega(k) = s\mid k \mid (1-\frac{\mid k\mid a}{2\pi}) \; , 
\label{eq.phs2}
\end{equation}
where $s=(\hbar/ma)\pi \sqrt{g/2}$ is the sound velocity.  This
coincides with the velocity of a compressional sound wave,  
$s=\sqrt{\partial p/\partial \rho}$, where $p=-\partial E/\partial L $ is
the pressure, $\rho=m/a$ is the density and E is given by Eq.\ (\ref{eq.egs}),
if $\beta$ is approximated as $\beta \simeq \sqrt{2g}$, which 
is justified for  $g\gg 1$.

The ground-state wavefunction in the harmonic approximation
$\Psi_h$ is such that
\begin{equation}
\mid \Psi_h \mid^2= {\mbox {const.}} \prod_k \exp{\{-\frac{m\omega(k)\mid
u_k\mid^2}{\hbar}\} } 
\; .
\label{eq.psh}
\end{equation}

Now we show that the exact wavefunction reduces to $\Psi_h$ 
for $ g \gg 1$. In this case $\beta \simeq
\sqrt{2g}\gg 1$ and it is justified to  apply the
harmonic approximation to the fictitious potential ${\cal V}(x)$ in
Eq.\ (\ref{eq.ps2}). Proceeding exactly as before we find
\begin{equation}
{\cal U} \equiv \frac{1}{2}\sum_{i\neq j} {\cal V}(x_i-x_j)
=\frac{1}{2}\sum_{n\neq m} {\cal V}(X_n-X_m) +\delta {\cal U}
\; ,
\label{eq.sca1}
\end{equation}
with
\begin{equation}
\delta {\cal U} = \frac{1}{2}\sum_k m   \Omega^2(k)\mid u_k \mid^2
\; , 
\label{eq.sca2}
\end{equation}
where $\Omega (k)$ is the fictitious potential phonon spectrum, 
given by Eq.\ (\ref{eq.phs}) with $V$ replaced by ${\cal V}$. 
Approximating ${\cal V}(x)$ by ${\cal V}(x)= -\ln{\mid  x \mid} $, the
sums in Eq.\ (\ref{eq.sca1}) can be performed exactly. The result is
\begin{equation}
m\Omega^2(k)=\frac{\pi}{a}\mid k \mid (1-\frac{\mid k\mid a}{2\pi})
\; .
\label{eq.omk}
\end{equation}

Substituting  Eq.\ (\ref{eq.omk}) in Eq.\ (\ref{eq.sca2}) and using
Eq.\ (\ref{eq.ps2}) it follows that the exact wavefuntcion 
coincides with  the harmonic approximation one in the limit $g\gg 1$.

\subsection{Structure Function and Correlation Function}
\label{sec.cor}

The structure function is defined  as
\begin{equation}
S(q)=\sum_{n} e^{iqX_n}\langle e^{iq(u_n- u_0)} \rangle
\; , 
\label{eq.sq}
\end{equation}
where  $\langle \rangle$ denotes the ground-state average. 

In the harmonic approximation, where the ground-state wavefunction,
Eq.\ (\ref{eq.psh}),  is  gaussian,
\begin{equation}
\langle e^{iq(u_n- u_0)} \rangle_h =e^{-\frac{1}{2}\langle \mid
q(u_n-u_0)\mid ^2\rangle_h} 
\; ,
\label{eq.dc}
\end{equation}
where  
\begin{equation}
\langle \mid q(u_n-u_0)\mid ^2\rangle_h =\frac{2q^2}{N}\sum_{k}\langle
\mid u_k \mid^2 \rangle_h (1-\cos{kX_n})
\; . 
\label{eq.qu2}
\end{equation}
It follows from Eq.\ (\ref{eq.psh}) that,
\begin{equation}
\langle \mid u_k \mid^2 \rangle_h=\frac{\hbar}{2m\omega(k)}
\; .
\label{eq.uk2}
\end{equation}
Using Eq.\ (\ref{eq.phs2}) we find that
\begin{equation}
\langle \mid q(u_n -u_0) \mid^2
\rangle_h = 2\eta^h(q)[\ln{(2\pi n) } + C - Ci(2\pi n)]
\; , 
\label{eq.qu2d}
\end{equation}
where $C=0.577$ is Euler's constant, $Ci(x)$ is the cosine-integral
function, defined as in Ref.\cite{abr}, and
\begin{equation}
\eta^h(q)=\frac{q^2 a^2}{\pi^2 \sqrt{2g}}
\; . 
\label{eq.eth}
\end{equation}

The above result shows that the
displacement correlation function, Eq.\ (\ref{eq.dc}), decays
algebraically with distance,  that is 
\begin{equation}
\langle e^{iq(u_n- u_0)} \rangle_h = F(n)\mid n\mid^{-\eta^h(q)} 
\; , 
\label{eq.dcf}
\end{equation}
where $F(n)=\exp{-\eta^h(q)[\ln{(2\pi ) } + C - Ci(2\pi n)]}$ is an
oscillating function of $n$.
From this result it follows, using Eq.\ (\ref{eq.sq}), that  for $q$ 
near a reciprocal lattice vector  $G_p=(2\pi/a)p$ ($p=$ integer) such
that $\eta^h(G_p)<1$,
$S(q)$, in the harmonic approximation, has power law singularities,
namely  
\begin{equation}
S_h(G_p+K)={\mbox { const.}}\mid K\mid^{\eta^h(G_p) -1}
\; , 
\label{eq.sgk}
\end{equation}
where $K\ll G_p$.

\section{Numerical Calculations}
\label{sec.nrs}

In this Section we discuss the Monte Carlo method used to calculate
ground-state averages of the type given by Eq.\ (\ref{eq.avt}) and
report the results obtained from it.

\subsection{Monte Carlo Method}
\label{sec.mcm}

Our numerical method consists of calculating averages from 
Eq.\ (\ref{eq.avt}) using
the traditional Monte Carlo method \cite{sma}. For a given $L$ we first
simulate the largest $\beta$ value. In this case the initial configuration is
an ordered chain of $N$ particles. New configurations are generated by
displacing particles, one at a time, in a sequential way along the
chain. The Metropolis algorithm is used to accept or reject
configurations. For smaller $\beta$ values we use as the initial
configuration the last one generated in the previous run. Typical runs
consist of 1000 MC steps per particle to equilibrate and 5000 MC steps
per particle to calculate averages. We simulate systems with fixed $a=50$
and with $N$ ranging from $N=100$ to $N=1000$. The use of a single $a$ value
is justified because in the CS model  the correlation functions depend only
on $a$ through $x/a$ or $qa$.

Our aim is to investigate by this method  the behavior of the
displacement correlation functions $\langle e^{iG_p(u_n- u_0)}
\rangle$, for $n\gg 1$. There are three possibilities.

\begin{description}

\item{i-} QLRPO: $
\langle e^{iG_p(u_n- u_0)} \rangle \rightarrow \mid n\mid^{-\eta(G_p)} \;$.

\item{ii-} True long-range positional order:
$\langle e^{iG_p(u_n- u_0)} \rangle \rightarrow \mbox {const.} 
\; .$
\item{iii-} Exponential decay: 
$\langle e^{iG_p(u_n- u_0)} \rangle \rightarrow e^{-n/\xi}
\; .$

\end{description}

For $L=\infty$, both  true long-range positional order and QLRPO 
with $\eta(G_p)<1$ lead  to  divergencies
in $S(G_p)$. For finite $L$ these singularities are replaced by  peaks
of finite height. In order to determine whether  peaks correspond to
singularities we perform the following finite size scalling  analysis of
the  data. For a given 
$\beta$  we compute  $S(G_p)$ for several values of $L$. Next we fit
this data to 
\begin{equation}
S(G_p)={\mbox { const.}}\mid L\mid^{1 - \eta(G_p) }
\; .
\label{eq.sgl}
\end{equation}
This dependence on $L$ arises as follows. If the system has
QLRPO with $\eta(G_p)<1$,   
Eq.\ (\ref{eq.sq})  predicts that $S(G_p)$ 
diverges with $L$ according to Eq.\ (\ref{eq.sgl}).
If true long-range positional order is present Eq.\ (\ref{eq.sq})  
 predicts $S(G_p)\sim L$  and 
Eq.\ (\ref{eq.sgl}) gives $\eta(G_p)=0$. 

This finite size scalling method cannot distinguish between QLRPO
 with $\eta(G_p)\ge 1$ and exponential decay. In both cases 
$S(G_p)$ becomes $L$-independent, because there is no singularity.
Thus, fitting the  $S(G_p)$ data to  Eq.\ (\ref{eq.sgl}) results 
in $\eta(G_p)=1$.  

In order to distinguish between these two possibilities
 we compute, for a given $L$, 
$\langle e^{iG_p(u_n- u_0)}\rangle$ and study its behavior as a
function of $n$.

\subsection{Results}
\label{sec.res}

\begin{figure}
\centerline{\epsfxsize= 6.0cm\epsfbox{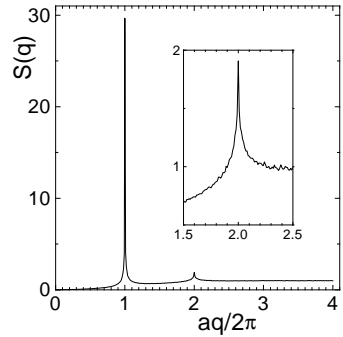}}
\caption{Structure function as a function of $q$ for $\beta=18$,
$L=5000$ and $N=100$. Inset shows in detail the peak at $q=G_2$.}
\label{fig.sq1}
\end{figure}

We find that  $S(q)$, has peaks at reciprocal
lattice vectors $q=G_p=2\pi p/a$, with $p=0,\pm 1,..,\pm p_{max}$.
The value of $p_{max}$ depends on  $\beta$ and increses with increasing
$\beta$. In  Fig.\ \ref {fig.sq1} we show  $S(q)$ for $\beta=18$,
where two peaks are observed at $q=G_1$ and $q=G_2$. As $\beta$
decreases, the peak at 
$q=G_2$ is no longer observed for $\beta \lesssim 12$. Below this value
there is only one peak at $q=G_1$. This peak disappears for a value of
$\beta < 4$. 

\begin{figure}
\centerline{\epsfxsize=6cm\epsfbox{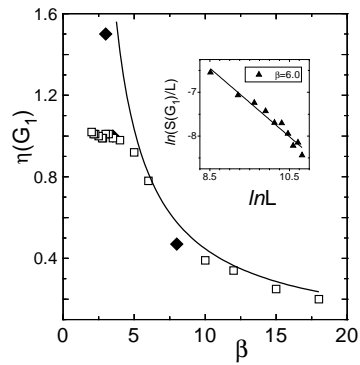}}
\caption{Exponent $\eta(G_1)$ as a function of $\beta$. Solid squares:
finite size scalling of MC data. Inset: typical plot used to obtain
$\eta(G_1)$. Continuous line: harmonic approximation.
Open squares: $\eta(G_1)$ obtained from the decay of 
$\langle e^{iG_1(u_n-u_0)}\rangle$ with $n$ for $L=5000$ and $N=100$.} 
\label{fig.eta}
\end{figure}

To identify whether or not  these peaks  correspond to singularities we 
carry out the finite size scalling analysis described above. This 
is illustrated in  Fig.\ \ref {fig.eta}. We find that the dependence of
the exponent $\eta(G_1)$  on $\beta$ is that shown in Fig.\ \ref {fig.eta}. 

\begin{figure}
\centerline{\epsfxsize=6cm\epsfbox{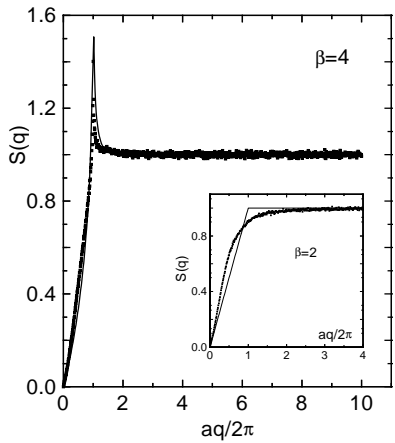}}
\caption{Structure function as a function of $q$. Comparison between MC
results for $L=5000$ and $N=100$ (solid squares) and exact results
(continuous line) for $\beta=4$ and $\beta=2$ (inset).}
\label{fig.sq2}
\end{figure}

We interpret the $\beta$-dependence of $\eta(G_1)$ shown in Fig.\ \ref
{fig.eta} as follows. The numerical estimates for $\eta(G_1)$ are in
good agreement with harmonic approximation predictions for $\beta > 5$. 
For $\beta=4$ Sutherland has obtained $S(q)$ exactly. A logarithmic
singularity occurs at $q=G_1$,   that is  $ S(q) \propto
\ln{\mid q-G_1\mid}$, rather than an algebraic one.  We find that  
at $\beta = 4$,  $\eta(G_1) = 0.98$ ( Fig.\ \ref {fig.sq2}). We
believe that this is 
consistent with Sutherland's exact result. The reason is that the
logarithm divergence leads to $S(G_1)\sim \ln{L}$ in a
finite system. It is well known that if this function is fitted to Eq.\
(\ref{eq.sgl}) on a log-log plot it leads to an exponent equal to
zero\cite{sta}, which  corresponds to   $\eta(G_1)=1$.   
Our estimate is, within the accuracy of our
simulations, consistent with that. For $\beta = 4$ our numerical
results for $S(q)$  agree well with the exact ones, as shown in Fig.\
\ref {fig.sq2} 
For $2<\beta<4$ our results fitted to Eq.\ (\ref{eq.sgl}) give
$\eta(G_1)\sim1$. As discussed in Sec.\ref{sec.mcm},  this only  indicates that
$S(q)$ has no singularity at $q=G_1$, which is consistent with either
QLRPO with $\eta(G_1)>1$ or with exponential decay.

\begin{figure}
\centerline{\epsfxsize= 6.0cm\epsfbox{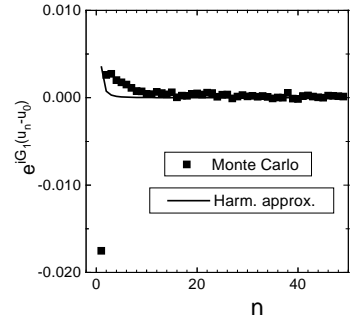}}
\caption{Displacement correlation function as a function of $n$. Data
points: MC data  for $L=5000$ and $N=100$. Continuous line: harmonic
approximation.}  
\label{fig.eig}
\end{figure}

In order to investigate which of these possibilities occur in the 
region $2<\beta<4$ we compute the displacement correlation function,
$\langle e^{iG_1(u_n- u_0)} \rangle$, as a function of $n$, for a given
$L$. We find that, as a result of strong quantum fluctuations, very
long MC runs are necessary in order to obtain reasonably accurate
correlation functions. Our results for $\beta=8.0,\; 3.5$ and $3$, shown in
Figs.2 and 4 require $10^6$ MC steps. For $\beta=3$ and $3.5$ we find that 
 $\langle e^{iG_1(u_n-u_0)} \rangle$   decays  with $n$ 
slower than the harmonic approximation predicts and even becomes
negative at $n=1$, as shown in  Fig.\ \ref {fig.eig}.  We also find
that for $\beta=3.0$ and $3.5$ it decays as $n^{-1.5}$ and $n^{-1.0}$,
respectively. 
We interpret this as evidence that QLRPO with $\eta(G_1)>1$ is also
present for $2<\beta<4$. Thus, there is no phase with exponential decay
of positional correlations in the CS model.

For $\beta=2$ Sutherland has also obtained $S(q)$ exactly. In  Fig.\
\ref {fig.sq2}  we
show that our results for $\beta=2$  agree reasonably well with the
exact ones. We attribute the rounding off near $q=G_1$ to finite size
effect.

We also calculate the exponent $\eta(G_2)$ by the finite size scalling
method for a few $\beta$ values. We find that, in the region where $S(G_2)$
has a singularity, $\eta(G_2)<1$ and its estimated value is in close
agreement with the harmonic
approximation one, $\eta^h(G_2)$. This approximation predicts that
$\eta^h(G_2)\leq 1$ for $\beta\geq 17$ ($g\geq 128$). We did not
attempt to estimate $\eta(G_2)$ outside the region where $S(G_2)$
has no singularity from the displacement correlation function $\langle
e^{iG_2(u_n- u_0)} \rangle$.

\section{Discussion}
\label{sec.dis}

The main conclusion of our numerical study is that the
ground-sate of the  CS model for $g>0$ or $2\leq \beta < \infty$
has  QLRPO characterized by an exponent that  
varies continuously with $\beta$. Our results also reveal that, as far
as  positional correlations are concerned, three distinct parameter
regions can be identified: i) A  semiclassical region  for $\beta > 5$
($g>7.5$) in which $\eta(G_1)\simeq \eta^h(G_1)$. As discussed in
Sec.\ref{sec.hap} 
this is expected for large $g$ where the exact CS ground-state
wavefunction coincides with  the harmonic approximation one. Our
results  show that for $g>7.5$ the CS model behaves semiclassically. 
ii) A region of moderate quantum fluctuations, for  $4<\beta < 5$
($4<g<7.5$), where $S(q)$ has one algebraic singularity at $q=G_1$ with
exponent  $\eta(G_1)<1$. In this region anharmonic quantum fluctuations
are important, leading to $\eta(G_1)<\eta^h(G_1)$. iii) A strong
quantum fluctuations region, for $2<\beta<4$,  where $S(q)$ has no
singularities, but QLRPO exists with  $\eta(G_1)>1$. As a result of
strong quantum fluctuations the displacement correlation function at
short distances becomes negative, in contrast to the harmonic
approximation that predicts  positive value for all $n$. At  large distances
it decays algebraically with distance but with an exponent such that
$\eta(G_1)\ll \eta^h(G_1)$.

It is instructive to estimate the amplitude of quantum fluctuations in
the boson positions using  a 'cage model'\cite{nelb}. This model
considers a single particle moving in the  potential well produced by
the other 
bosons, assumed fixed  in their classical positions. In this case we
find that, in the harmonic approximation, the boson zero-point  mean-square 
displacement from its classical equilibrium position is   $\langle
u^2 \rangle/a^2=0.55/\sqrt{g}$.  
According  to the discussion above, the semiclassical region
corresponds to $g>7.5$, for which $\sqrt{\langle u^2
\rangle/a^2}<0.45$. We propose that this last result be adopted
as a kind of 'Lindemann criterion' \cite{dp} to estimate the range of
validity of the harmonic approximation for 1D interacting boson models
in general.  

An important conclusion that can be drawn from these results is that, as
far as QLRPO is concerned, the harmonic approximation gives a correct 
qualitative picture for the behavior of the CS model ground-state.
The differences that we find between the exact result and this
approximation, although large for $\beta<5$, are quantitative rather than
qualitative. The fact that this conclusion is reached  in an exactly
soluble model suggests that for other 1D boson models the harmonic
approximation predictions for QLRPO  are correct, at least
qualitatively, well beyond the semiclassical region, as long  as a
phase transition does not take place. 

One possibility  that cannot be studied by our method is the existence
of a superfluid phase. We believe that this phase can be ruled out in
the CS model because the bosons are impenetrable. In 1D this excludes the
possibility of a superfluid phase. The reason is that if one constructs the
path-integral representation for the ground-state partition function,
the superfluid density is related to the boson world lines
winding number fluctuations\cite{cp}. In 1D  these fluctuations require that 
the boson world lines cross each other, which is not possible if 
the bosons are impenetrable. From this and from our numerical results
we conclude that the ground-state of the  repulsive CS model has only
one normal phase with QLRPO.  We believe that a large class of 1D
models for impenetrable bosons interacting through a repulsive
potential falling off so fast with distance to exclude
true long-range positional order has a similar ground-state phase diagram.

\acknowledgments
It is a pleasure to thank Prof. E. Mucciolo for stimulating  conversations.
This work was supported in part by FINEP/Brazil,
CNPq-Bras\'{\i}lia/Brazil and Funda\c{c}\~ao Universit\'aria Jos\'e
Bonif\'acio.

\end{multicols}


\begin{references} 

\bibitem[*]{em} e-mail:izzo@if.ufrj.br


\bibitem{mer} N.D. Mermin, Phys. Rev.  {\bf 176}, 250 (1968).

\bibitem{nel} D.R. Nelson {\em Phase Transitions and Critical Phenomena
Vol. 7}, edited C. Domb and M.S. Green, (Academic Press, London, 1983)
 
\bibitem{tch} J. Tobochnik and G. V. Chester, Phys. Rev. B {\bf 25},
6778 (1982). 

\bibitem{cs} B. Sutherland, Phys. Rev. A  {\bf 4}, 2019 (1971); J.
Math. Phys. {\bf 12}, 246 (1971).


\bibitem{strh} V. Ya. Krivnov and A.A. Ovchinnikov, Zh. Eksp. Teor.
Fiz.  {\bf 82}, 271 (1982) [ Sov. Phys. JETP {\bf 55}, 162 (1982)]; 
A. V. Zabrodin and A.A. Ovchinnikov, Zh. Eksp. Teor.
Fiz.  {\bf 90}, 2260 (1986) [ Sov. Phys. JETP {\bf 63}, 1326 (1986)]; 
P.J. Forrester, J. Stat. Phys. {\bf 72}, 39 (1993); D. Sen and R.K.
Bhaduri, preprint cond-mat 9705279

\bibitem{sma} See, e.g., {\em Monte Carlo Methods in Statistical
Physics}, edited  by  K. Binder (Springer, Berlin, 1979).

\bibitem{dp} D. Pines {\em Elementary Excitations in Solids}
(Benjamin, N.Y., 1963).  

\bibitem{abr} M. Abramowitz I.A. Segun, {\em Handbook of Mathematical
Functions}  (Dover, N.Y., 1965).

\bibitem{sta} H.E. Stanley  {\em Introduction to Phase Transitions}
(Oxford, Oxford, 1971).  

\bibitem{nelb} E. Frey, D.R. Nelson and D.S. Fisher, Phys. Rev. B {\bf 49},
9723 (1994). 

\bibitem{cp} D.M. Ceperley and E.L. Pollock, Phys. Rev. B {\bf 39},
2084 (1989). 


\end{references}
\end{document}